# On Nonsymmetric Nonparametric Measures of Dependence

*Hui Li*[1]

Based on recent progress in research on copula based dependence measures, we review the original Rényi's axioms on symmetric measures and propose a new set of axioms that applies to nonsymmetric measures. We show that nonsymmetric measures can actually better characterize the relationship between a pair of random variables including both independence and complete dependence. The new measures also satisfy the Data Processing Inequality (DPI) on the $*$ product on copulas, which leads to nice features including the invariance of dependence measure under bijective transformation on one of the random variables. The issues with symmetric measures are also clarified.

1. Introduction

With the advance of modern science and technology, the amount of data generated is growing exponentially. To model the underlying relationship between different variables behind the data, it is very important to be able to measure the strength of the statistical dependence between them, which will be the starting point toward more structural analysis. For this purpose, it would be convenient to have a scalar value representing and ranking the strength of dependence between any two random variables. The scalar measure should capture the major dependence relationships between them from the extreme of independence to that of complete dependence. This kind of measure has been a keen research topic in statistics and related sciences for a long time, with applications to fields including information communications, data mining, economics, biomedical research and artificial intelligence.

With the amount of data available, it would also be very helpful if the scalar measure is nonparametric as the complex relationships may not be measured in the right units and be easily parameterized. Although previous research has mainly focused on symmetric measures of dependence, it is obvious that dependence relationship is in general nonsymmetric, especially for functional relationships. Existing symmetric measures have the problem of taking maximum value on a much wider range of dependence relationships besides complete dependence. It is already known that some nonsymmetric measures take maximum value only on complete dependence (Dette, Siburg and Stoimanov, 2010; Trutschnig, 2011). So a nonsymmetric measure would be more realistic and reveals more information. Another popular research topic is on the copula (see for example Nelson (2006)), which captures all the dependence information among continuous random variables irrespective of marginal distributions thus nonparametric. Hence a copula-based dependence measure would be natural. We will see that, instead of copula itself or its density, which normally lead to symmetric dependence measures, it is the first-order partial derivatives of copula that can be used to construct nonsymmetric nonparametric measures.

The current paper will discuss a new class of nonsymmetric nonparametric measures of dependence. Section 2 reviews the axioms of symmetric nonparametric dependence measure proposed by Rényi (1959) and presents our new axioms for nonsymmetric measure. Section 3 reviews the basic properties of copula which forms the foundation for the study of the new measures. Section 4 presents the new class of nonsymmetric dependence measures and shows that they satisfy the new axioms. It also discusses the Data Processing Inequality (DPI) on the ∗ product on copulas and its implication on the properties of the dependence measures. Section 5 gives an example and verifies the properties of the new measures. Section 6 concludes the paper. As the new dependence measures are distance-like functions, we present entropy-like measures in the appendix for completeness.

## 2. Axioms of dependence measure

In 1959, Rényi (1959) introduced a set of axioms as the criteria of a symmetric nonparametric measure of dependence $R(X,Y)$ for two random variables $X,Y$ on a common probability space.

a) $R(X,Y)$ is defined for all non-constant random variables $X,Y$;
b) $R(X,Y) = R(Y,X)$;
c) $0 \leq R(X,Y) \leq 1$;
d) $R(X,Y) = 0$ if and only if $X,Y$ are independent;
e) $R(X,Y) = 1$ if either $Y = f(X)$ or $X = g(Y)$ almost surely for some Borel-measurable functions $f, g$;
f) If $f$ and $g$ are Borel-measurable bijections on $\mathbb{R}$, then $R(f(X), g(Y)) = R(X,Y)$;
g) If $X,Y$ are jointly normal with correlation coefficient ρ, then $R(X,Y) = |\rho|$.

The first known measure of dependence that satisfies all the axioms is the maximal correlation coefficient introduced by Gebelein (1941),

$$\tilde{\rho}(X,Y) = \sup_{f,g} \rho(f(X), g(Y)) \qquad (1)$$

where $f, g$ are Borel measurable functions and $\rho$ is the linear correlation. It is hard to calculate and equals to one too often. Another well-known measure is the information coefficient of correlation (Linfoot, 1957),

$$MIcor(X,Y) = \sqrt{1 - e^{-2MI(X,Y)}} \tag{2}$$

which is a scaled version of Shannon's mutual information (Shannon and Weaver, 1949; Cover and Thomas, 1991)

$$MI(X,Y) = \iint f(x,y) \log\left(\frac{f(x,y)}{f(x)f(y)}\right) dxdy \tag{3}$$

where $f(x,y)$ is the density of the joint distribution and $f(x), f(y)$ are densities of the marginal distributions. It is used in a lot of areas but equals to 1 whenever $f(x,y)$ has singular value.

Rényi's condition e) is not very strong as $Y = f(X)$ or $X = g(Y)$ is sufficient but not necessary. The known measures that satisfy all Rényi's conditions are equal to one even when the relation may not be complete dependence. On the other hand, condition e) is also deemed to be too strong. As Kimeldorf and Sampson (1978) found out, sequences of pairs of completely dependent variables may not converge to a pair of completely dependent variables, but sequences of pairs of monotone dependent variables will converge to a pair of monotone dependent variables. This led to the exploration of other dependence measures.

Schweizer and Wolff (1981) introduced a set of relaxed conditions where Borel-measurable functions in e) and f) are replaced with strictly monotonic functions as necessary and sufficient condition, the random variables are now continuously distributed and $R(X,Y)$ is strictly increasing function of $|\rho|$ for normal $X,Y$, which is an equivalent form of condition g). Note that mutual information satisfies this equivalent form of g) and Linfoot's information coefficient of correlation maps it to $|\rho|$. For continuous random variables, the dependence can be characterized by a unique copula function (Sklar, 1959) which is independent of the marginal distributions and has good properties under strictly monotonic transformations. If $F_{X,Y}$ is the joint distribution function of $X$ and $Y$ with marginal distributions $F_X$ and $F_Y$, then there exists a unique copula function $C$ on $[0,1] \times [0,1]$ such that

$$F_{X,Y}(x,y) = C(F_X(x), F_Y(y)) \tag{4}$$

A nonparametric measure of dependence should be function of the copula alone. Specifically, Schweizer and Wolff (1981) studied the suitably normalized $L^1$, $L^2$ and $L^\infty$ distances between any copula and the independence copula as measures of dependence and proved that they satisfy the relaxed conditions:

$$\sigma(X,Y) = 12 \int_0^1 \int_0^1 |C(u,v) - uv| dudv \tag{5}$$

$$\gamma(X,Y) = \left(90 \int_0^1 \int_0^1 [C(u,v) - uv]^2 \, du\, dv\right)^{1/2} \tag{6}$$

$$\kappa(X,Y) = 4 \sup_{u,v \in [0,1]} |C(u,v) - uv| \tag{7}$$

Recently, Siburg and Stoimenov (2009) introduced a measure of mutual complete dependence defined through the Sobolev norm on copulas. It is again defined only on continuous random variables and satisfies conditions b) – d) but with e) and f) modified as follows e') $R(X,Y) = 1$ if and only if $Y = f(X)$ and $X = g(Y)$ almost surely for some Borel-measurable bijections $f, g$;

f') If $f$ and $g$ are strictly monotonic functions, then $R(f(X), g(Y)) = R(X,Y)$;

Note that f') is the same requirement as Schweizer and Wolff (1981), but e') requires that $R(X,Y) = 1$ only if $X, Y$ are mutually completely dependent. A later article by Ruankong, Santiwipanont and Sumetkijakan (2012) constructed a new measure of dependence based on the measure of Siburg and Stoimenv (2009) and showed that it satisfies the original Rényi conditions b) – f). The drawbacks of this new measure are that it might be hard to calculate and equals to one even when $X$ or $Y$ is not a function of the other. Therefore, none of the afore mentioned measures are satisfactory.

It is well-known that although independence is a symmetric property, complete dependence is not. If $X$ is a function of $Y$, then $X$ is completely dependent on $Y$ but $Y$ need not to be completely dependent on $X$ unless the function is a bijection. So Rényi's condition b) is somewhat unintuitive. Recently, nonsymmetric measures of dependence have started to attract some attentions as new research on properties of copula naturally leads to them, see Dette, Siburg and Stoimenov (2010), Trutschnig (2011). Although these new measures are interesting, a systematic study along the similar lines of Rényi's axioms has not been carried out. It is the purpose of the present paper to initiate the research in that direction. Specifically, we assume $R(X,Y)$ measures the degree of dependence of $Y$ on $X$ and satisfies the following conditions:

a") $R(X,Y)$ is defined for all non-constant continuous random variables $X, Y$;

b") $R(X,Y)$ may not be equal to $R(Y,X)$;

c") $0 \leq R(X,Y) \leq 1$;

d") $R(X,Y) = 0$ if and only if $X, Y$ are independent;

e") $R(X,Y) = 1$ if and only if $Y = f(X)$ almost surely for a Borel-measurable function $f$;

f") If $g$ is a Borel-measurable bijection on $\mathbb{R}$, then $R(g(X), Y) = R(X,Y)$;

Condition a") restrict the random variables to continuous ones such that the copula between them is uniquely defined. Condition b") specifies that the dependence measure can be nonsymmetric. But if a copula is symmetric or $C(u, v) = C(v, u)$, then $R(X, Y) = R(Y, X)$. Conditions c") and d") are the same as Rényi's conditions c) and d) as independence is a symmetric property. Condition e") is more explicit about the nonsymmetric nature of dependence and is stronger as $R(X, Y) = 1$ happens if and only if $Y = f(X)$ almost surely. As a nonsymmetric measure, condition f'') only requires the measure to be invariant under bijective transformation on $X$. As we will see later, certain measures may also be invariant under strictly monotonic transformation on $Y$, but it is not universal. Note that condition g) is omitted as normal correlation ρ is symmetric and may not be relevant here, but it is perfectly possible that the dependence measure is an increasing function of $|\rho|$ for normal $X, Y$.

## 3. Basic properties of bivariate Copula

Let $I$ denote the closed unit interval $[0,1]$ and $I^2$ the closed unit square $[0,1] \times [0,1]$.

DEFINITION 3.1. A bivariate copula is a function $C: I^2 \to I$ that satisfies the following conditions:

(i) $C(u, 0) = C(0, v) = 0$ for all $u, v \in I$.

(ii) $C(u, 1) = u$ and $C(1, v) = v$ for all $u, v \in I$.

(iii) $C(u_2, v_2) - C(u_2, v_1) - C(u_1, v_2) + C(u_1, v_1) \geq 0$ for all $u_1, v_1, u_2, v_2 \in I$ such that $u_1 \leq u_2$ and $v_1 \leq v_2$.

Copulas are of interest because they link joint distributions to marginal distributions. Sklar (1959) showed that, for any real random variables $X, Y$ with continuous cumulative distribution functions $F_X$, $F_Y$ and joint distribution function $F_{X,Y}$, there is a unique copula $C$ such that

$$F_{X,Y}(x, y) = C(F_X(x), F_Y(y))$$

Let $\mathbb{C}$ be the set of all bivariate copulas. Denote the partial derivative of $C \in \mathbb{C}$ with respect to the $i$-th variable as $\partial_i C$ where $i \in \{1, 2\}$. We list the following key properties of copula; for a proof, see for example Nelson (2006).

PROPOSITION 3.2. (i) $C$ is nondecreasing in each argument.

(ii) $\mathbb{C}$ is closed under convex combinations, i.e., if $A, B \in \mathbb{C}$ and $a, b \in I$ with $a + b = 1$, then $aA + bB \in \mathbb{C}$.

(ii) $|C(u_2, v_2) - C(u_1, v_1)| \leq |u_2 - u_1| + |v_2 - v_1|$, hence $C$ is Lipschitz and uniformly continuous.

(iii) For $i \in \{1,2\}$, $\partial_i C$ exists almost everywhere on $I^2$ with $0 \leq \partial_i C \leq 1$.

(iv) The functions $v \to \partial_1 C(u,v)$ and $u \to \partial_2 C(u,v)$ are defined and nondecreasing almost everywhere on $I$.

Note that condition (ii) in Definition 3.1 means $\partial_1 C(u,1) = 1$ and $\partial_2 C(1,v) = 1$.

Any copula $C$ can be decomposed into the sum of an absolutely continuous part

$$A_C(u,v) = \int_0^u \int_0^v \partial_1 \partial_2 C(s,t) ds dt \tag{8}$$

and a singular part with support on a zero-measure set

$$S_C(u,v) = C(u,v) - A_C(u,v) \tag{9}$$

If a copula does not have a singular part, it is absolutely continuous such that $\partial_1 \partial_2 C(u,v)$ and $\partial_2 \partial_1 C(u,v)$ exist almost everywhere, are bounded and integrable, and are equal almost everywhere. The absolutely continuous copulas are dense in the set of all copulas.

There are three well-known copulas

$$W(u,v) = \max(u + v - 1, 0) \tag{10}$$

$$M(u,v) = \min(u,v) \tag{11}$$

$$\Pi(u,v) = uv \tag{12}$$

$W$ and $M$ are called the Fréchet-Hoeffding lower and upper bounds as for any copula $C$,

$$W(u,v) \leq C(u,v) \leq M(u,v) \tag{13}$$

$\Pi$ is the independent or product copula. $M$ ($W$) is a copula of $X$ and $Y$ if and only if $Y$ is almost surely a strictly increasing (decreasing) Borel-measurable function of $X$, where $\Pi$ is a copula of $X$ and $Y$ if and only if $X, Y$ are independent. $W$ and $M$ are singular copulas while $\Pi$ is absolutely continuous with density 1.

Next we define the $*$ product, which was introduced by Darsow, Nguyen and Olsen (1992).

DEFINITION 3.3 For any two copula functions A and B, the $*$ product is defined as

$$(A * B)(u,v) = \int_0^1 \partial_2 A(u,t) \cdot \partial_1 B(t,v) dt \tag{14}$$

It is easy to show that $A * B$ is again a copula function, see Darsow, Nguyen and Olsen (1992). Actually the independent copula $\Pi$ is the null element and the upper copula $M$ is the unit element for the $*$ product, such that, for any copula $C$,

$$C * \Pi = \Pi * C = \Pi \tag{15}$$

$$C * M = M * C = C \tag{16}$$

It can be shown that the $*$ product is associative, thus the set $\mathbb{C}$ becomes a monoid with a null element and a unit element. Although $\mathbb{C}$ is not a group, some of its elements do have inverses.

DEFINITION 3.4 Let $C \in \mathbb{C}$.

(i) $C$ is called left invertible if there a copula A such that $A * C = M$.

(ii) $C$ is called right invertible if there a copula B such that $C * B = M$.

(iii) $C$ is invertible if it is both left and right invertible.

It can be shown that the left or right inverse of a copula is unique and corresponds to the transposed copula $C^T(u, v) = C(v, u)$, see e.g. Darsow, Nguyen and Olsen (1992), which also proved the necessary and sufficient condition for copula invertibility.

PROPOSITION 3.5 Let $C \in \mathbb{C}$.

(i) $C$ is left invertible if and only if for each $v \in I$, $\partial_1 C(u, v) \in \{0,1\}$ for almost all $u \in I$.

(ii) $C$ as the copula between random variables $X$ and $Y$ is left invertible if and only if there is a Borel function $f$ such that $Y = f(X)$ almost surely.

(iii) $C$ is right invertible if and only if for each $u \in I$, $\partial_2 C(u, v) \in \{0,1\}$ for almost all $v \in I$.

(iv) $C$ as the copula between random variables $X$ and $Y$ is right invertible if and only if there is a Borel function $g$ such that $X = g(Y)$ almost surely.

A copula invertible on one side needs not to be invertible on the other side, which is another way to express the nonsymmetric nature of dependence. Invertible copulas imply mutual complete dependence between two random variables.

The following proposition comes from Theorem 2.3 in Darsow, Nguyen and Olsen (1992).

PROPOSITION 3.6 Consider copulas $A_n, A, B$ such that $A_n \to A$. Then $A_n * B \to A * B$ and $B * A_n \to B * A$.

## 4. A new class of nonsymmetric nonparametric measures

The nonsymmetric measures defined in Dette, Siburg and Stoimenov (2010) and Trutschnig (2011) are based on first order partial derivative of the copula function and have the distance-like form

$$\tau(C) = \int_0^1 \int_0^1 \varphi(\partial_1(C(u,v) - \Pi(u,v)))dudv \tag{17}$$

where $\Pi(u,v) = uv$ is the independent copula. So the measure is in general a normalized distance between any copula and the independent copula, either originated from the Sobolev norm (Siburg and Stoimanov, 2009) or from the induced metric on Markov operators (Trutschnig, 2011). Specifically, the two existing measures are

$$\tau_1(C) = 3\int_0^1 \int_0^1 |\partial_1(C(u,v) - \Pi(u,v))|dudv \tag{18}$$

and

$$\tau_2^2(C) = 6\int_0^1 \int_0^1 (\partial_1(C(u,v) - \Pi(u,v)))^2 dudv \tag{19}$$

The conditions a") – e") have already been verified in Dette, Siburg and Stoimenov (2010) and Trutschnig (2011) for $\tau_1$ and $\tau_2$, but condition f") has not been verified. The proof of condition f") will be discussed below.

PROPOSITION 4.1 If random variables $X$ and $Z$ are conditionally independent given $Y$, then $C_{XZ} = C_{XY} * C_{YZ}$.

This was shown in Darsow, Nguyen and Olsen (1992), p610. We can say $X, Y, Z$ form a Markov chain. $Z$ is less dependent on $X$ than on $Y$ as the dependence of $Z$ on $X$ is through $Y$. This should be reflected in the dependence measure.

PROPOSITION 4.2 For the general form of nonsymmetric dependence measure in (17), if $\varphi$ is a convex function and is not directly dependent on $u$, then $\tau(C_{XZ}) \leq \tau(C_{YZ})$, if $X, Y, Z$ form a Markov chain.

PROOF: It suffices to consider that $C_{XY}$ is absolutely continuous according to Proposition 3.6 and the fact that absolutely continuous copulas are dense in the set of all copulas. Using Jensen's inequality,

$$\tau(C_{XZ}) = \tau(C_{XY} * C_{YZ})$$

$$= \int_0^1 \int_0^1 \varphi(\partial_1((C_{XY} * C_{YZ})(u,v) - \Pi(u,v)))dudv$$

$$= \int_0^1 \int_0^1 \varphi\big(\partial_1((C_{XY} * (C_{YZ} - \Pi))(u,v)\big)dudv$$

$$= \int_0^1 \int_0^1 \varphi\left(\int_0^1 \partial_1\partial_2 C_{XY}(u,t) \cdot \partial_1(C_{YZ} - \Pi)(t,v)dt\right)dudv$$

$$\leq \int_0^1 \int_0^1 \left(\int_0^1 \partial_1\partial_2 C_{XY}(u,t) \cdot \varphi(\partial_1(C_{YZ} - \Pi)(t,v))dt\right)dudv$$

$$= \int_0^1 \left(\int_0^1 (\int_0^1 \partial_1\partial_2 C_{XY}(u,t)\, du) \cdot \varphi(\partial_1(C_{YZ} - \Pi)(t,v))dt\right)dv$$

$$= \int_0^1 \int_0^1 \varphi(\partial_1(C_{YZ} - \Pi)(t,v)) dt\, dv$$

$$= \tau(C_{YZ}) \tag{20}$$

where we have used the following

$$\int_0^1 \partial_1 \partial_2 C_{XY}(u,t)\, du = \partial_2 C_{XY}(1,t) = \partial_2 t = 1 \tag{21}$$

and

$$\int_0^1 \partial_1 \partial_2 C_{XY}(u,t)\, dt = \partial_1 C_{XY}(u,1) = \partial_1 u = 1 \tag{22}$$

If $\varphi$ is strictly convex, then the equal sign holds if $\partial_1(C_{YZ} - \Pi)(t,v)$ is almost constant in t with respect to the measure defined by the density $\partial_1 \partial_2 C_{XY}(u,t)$ on $t \in [0,1]$ for almost all $u, v \in [0,1]$. □

This is related to the Data Processing Inequality (DPI) in information theory, see, e. g., Chapter 2 of Cover and Thomas (1991). DPI says that if $X, Y, Z$ form a Markov chain, then $I(X,Y) \geq I(X,Z)$ for Shannon's mutual information. It implies that no processing of $Y$ can increase the information that $Y$ contains about $X$. A more general form of DPI for symmetric dependence measure was discussed in Kinney and Atwal (2014). Condition f'') is similar to the self-equitable property defined in their paper except that the measure here is nonsymmetric.

PROPOSITION 4.3  If $f$ is a Borel-measurable bijection on $\mathbb{R}$, then $\tau(C_{f(X)Y}) = \tau(C_{XY})$.

PROOF: As $f$ is a bijective mapping, $X, f(X), Y$ form a Markov chain. Thus $\tau(C_{XY}) \leq \tau(C_{f(X)Y})$. On the other hand, for any mapping $f$, $f(X), X, Y$ also form a Markov chain, which implies $\tau(C_{f(X)Y}) \leq \tau(C_{XY})$. Therefore $\tau(C_{f(X)Y}) = \tau(C_{XY})$. □

PROPOSITION 4.4  If $g$ is a strictly monotonic transformation on $\mathbb{R}$, then $\tau(C_{Xg(Y)}) = \tau(C_{XY})$.

PROOF: If $g$ is strictly increasing, then $C_{Xg(Y)} = C_{XY}$. If $g$ is strictly decreasing, $C_{Xg(Y)}(u,v) = u - C_{XY}(u, 1-v)$, see e.g. Nelson (2006). Therefore, $(C_{Xg(Y)} - \Pi)(u,v) = (\Pi - C_{XY})(u, 1-v)$, which, by change of variable, leads to $\tau(C_{Xg(Y)}) = \tau(C_{XY})$ as long as the measure is symmetric to $C_{XY} - \Pi$ and $\Pi - C_{XY}$, which is generally true for distance-like measures. □

Therefore, for the general form of dependence measure in Equation (17) with convex function $\varphi$, we have proven a stronger condition than f'').

> f''') If $f$ is a Borel-measurable bijection and $g$ is a strictly monotonic transformation on $\mathbb{R}$, then $R(f(X), g(Y)) = R(X,Y)$.

It will be shown in the Appendix that other forms of nonsymmetric dependence measures exist that do not satisfy Proposition 4.4, which is why it is not listed in condition f").

Next we discuss some general properties of dependence measures that satisfy the DPI condition. For nonsymmetric measures, the DPI condition can be defined as

$$\tau(A*B) \leq \tau(B) \quad \text{or} \quad \tau(A*B) \leq \tau(A) \tag{23}$$

where $A, B$ are any copula function. Let us focus on the first one, which measures the dependence of $Y$ on $X$ for copula $C_{XY}$.

PROPOSITION 4.5 If $\tau(A*B) \leq \tau(B)$ holds, then $\tau(\Pi) \leq \tau(C) \leq \tau(M)$. If a copula $C$ is left invertible with respect to the $*$ product, then $\tau(C) = \tau(M)$. Conversely, for measures defined in Equation (17) with strict convex function $\varphi$, if $\tau(C) = \tau(M)$, then C is left invertible.

PROOF: We have $C*M = C$, which leads to $\tau(C) \leq \tau(M)$. Besides, $\Pi * C = \Pi$ leads to $\tau(\Pi) \leq \tau(C)$. Thus $\tau(\Pi) \leq \tau(C) \leq \tau(M)$. If C is left invertible, then $C^T * C = M$. This leads to $\tau(C) \geq \tau(M)$. Therefore $\tau(C) = \tau(M)$.

If $\tau(C) = \tau(M)$ for a copula C, then $\tau(C*M) = \tau(M)$. Based on the condition for equality in Equation (20), $\partial_1 M(t,v) - v$ is almost constant in t with respect to the measure defined by the density $\partial_1 \partial_2 C(u,t)$ on $t \in [0,1]$ for almost all $u, v \in [0,1]$. As $\partial_1 M(t,v) - v$ takes value $1-v$ when $t < v$ and $-v$ when $t > v$, the measure defined by $\partial_1 \partial_2 C(u,t)$ has to be singular and have mass at only one point on $t \in [0,1]$ for almost all u. This means C has support on the graph of a function $f(u)$ almost surely, so it is almost surely equal to $C_{u,f(u)}$ which is left invertible. □

If $\tau(\Pi)$ and $\tau(M)$ are finite, the range for $\tau(C)$ can always be normalized to $[0,1]$ such that $\tau(\Pi) = 0$ and $\tau(M) = 1$. For example, for dependence measures of the form

$$\tau_\alpha(C) = \left(\int_0^1 \int_0^1 |\partial_1(C(u,v) - \Pi(u,v))|^\alpha du dv\right)^{1/\alpha} \qquad \alpha \geq 1, \tag{24}$$

the normalization constant can be calculated as $k_\alpha = \frac{1}{\tau_\alpha(M)} = \left(\frac{(\alpha+1)(\alpha+2)}{2}\right)^{1/\alpha}$.

Note that a right invertible copula $C$ may not have $\tau(C) = \tau(M)$ as will be seen in the example in next section. However, its transpose will be left invertible and has $\tau(C^T) = \tau(M)$. This suggests that, to understand the dependence relation between two random variables, we may need to calculate both $\tau(C)$ and $\tau(C^T)$ in order to know which variable is completely dependent on the other. But the added calculation will provide us with more information about the dependence relation. The symmetric measures defined previously cannot provide this extra information.

The set of left invertible copulas $\mathcal{L}$ is closed under the $*$ product, thus forming a sub-monoid without null element. The dependence measure has the same value $\tau(M)$ on $\mathcal{L}$. The condition e") requires that $\mathcal{L}$ is the largest set to have this property, which is proven in Proposition 4.5.

PROPOSITION 4.6 Define the left coset of a copula $C$ as $\mathcal{L} * C = \{L * C, L \in \mathcal{L}\}$. Then the dependence measure has the same value $\tau(C)$ on all elements of $\mathcal{L} * C$.

PROOF: This is easy to show if we notice that $L^T * L * C = C$. Thus

$$\tau(C) = \tau(L^T * L * C) \leq \tau(L * C) \leq \tau(C).$$

This means that $\tau(L * C) = \tau(C)$ for any left invertible copula $L$. □

Similar results hold for measures satisfying the second DPI condition $\tau(A * B) \leq \tau(A)$ in Equation (23) which measures the dependence of $X$ on $Y$ for copula $C_{XY}$. This kind of measures can be obtained by transposing the copula in Equation (17) or by changing $\partial_1 C(u, v)$ to $\partial_2 C(u, v)$ in Equation (17). In this case the dependence measure will have the same value on the right coset $C * \mathcal{R}$ where $\mathcal{R}$ is the set of all right invertible copulas. A similar version of condition f'') is satisfied where the dependence measure is unchanged if $Y$ is transformed by a Borel measurable bijection.

For symmetric dependence measures which satisfies $\tau_S(C^T) = \tau_S(C)$, if one of the DPI conditions holds, then both of the DPI conditions will hold.

PROPOSITION 4.7 For a symmetric dependence measure $\tau_S$, if $\tau_S(A * B) \leq \tau_S(B)$, then $\tau_S(A * B) \leq \tau_S(A)$.

PROOF: If $C = A * B$, then $C^T = B^T * A^T$. Thus,

$$\tau_S(A * B) = \tau_S(C) = \tau_S(C^T) = \tau_S(B^T * A^T) \leq \tau_S(A^T) = \tau_S(A). \qquad \square$$

PROPOSITION 4.7 A symmetric dependence measure has maximum value $\tau_S(M)$ on $\mathcal{L} * \mathcal{R} = \{L * R, L \in \mathcal{L}, R \in \mathcal{R}\}$.

PROOF: If $D$ is a right invertible copula, then $\tau_S(D) = \tau_S(D^T) = \tau_S(M)$. If $C$ is left invertible, then $C^T * (C * D) = D$, and $\tau_S(D) = \tau_S(C^T * (C * D)) \leq \tau_S(C * D) \leq \tau_S(D)$. Therefore, $\tau_S(C * D) = \tau_S(D)$ and the dependence measure has the same value $\tau_S(M)$ on $\mathcal{L} * \mathcal{R}$. □

This property is already known for certain symmetric measures, see Corollary 5.5 of Ruankong, Santiwipanont and Sumetkijakan (2012). It is the key issue for symmetric dependence measures as copulas in $\mathcal{L} * \mathcal{R}$ may not mean complete dependence for either $Y$ on $X$ or $X$ on $Y$.

PROPOSITION 4.7 A symmetric dependence measure has the same value $\tau_S(C)$ on the double coset $\mathcal{L} * C * \mathcal{R} = \{L * C * R, L \in \mathcal{L}, R \in \mathcal{R}\}$ for any copula $C$.

PROOF: For a general copula $C$, as $L^T * (L * C * R) * R^T = C$, we have

$$\tau_S(C) \leq \tau_S(L^T * (L * C * R)) \leq \tau_S(L * C * R) \leq \tau_S(C * R) \leq \tau_S(C). \qquad \square$$

The symmetric measures are usually functions of the copula itself or its density, such as those discussed by Schweizer and Wolff (1981). Other examples of symmetric dependence measures include Rényi's mutual information (Rényi, 1961),

$$MI_\alpha(C) = \frac{1}{\alpha-1} \log \left[ \int_0^1 \int_0^1 c^\alpha(u,v) du dv \right], \quad \alpha > 0 \text{ and } \alpha \neq 1 \tag{25}$$

Tsallis entropy (Tsallis, 1988)

$$\Delta_\alpha(C) = \frac{1}{\alpha-1} \left[ 1 - \int_0^1 \int_0^1 c^\alpha(u,v) du dv \right], \quad \alpha > 0 \text{ and } \alpha \neq 1 \tag{26}$$

and the Copula-Distance

$$CD_\alpha = \int_0^1 \int_0^1 |c(u,v) - 1|^\alpha du dv, \quad \alpha \geq 1 \tag{27}$$

where $c(u,v) = \partial_1 \partial_2 C(u,v)$ is the copula density and 1 is the density of the independent copula, see Ding and Li (2013). Note that the popular Shannon's mutual information

$$MI(C) = \int_0^1 \int_0^1 c(u,v) \cdot \log(c(u,v)) du dv \tag{28}$$

can also be included as the limit $\alpha \to 1$ of both Rényi's mutual information and Tsallis entropy.

The $*$ product has a simpler form on copula densities:

$$c(u,v) = a(u,v) * b(u,v) = \int_0^1 a(u,t) \cdot b(t,v) dt \tag{29}$$

where $a(u,v)$ and $b(u,v)$ are densities of copulas $A$ and $B$.

The DPI property can be proven similarly for dependence measures as convex functions of the copula density, see also Kinney and Atwal (2014). Let the dependence measure be a function of the copula density as follows:

$$\tau(C) = \int_0^1 \int_0^1 \varphi(c(u,v)) du dv \tag{30}$$

where $\varphi$ is a convex function and does not depend on $u$ directly. We have

$$\tau(A * B) = \int_0^1 \int_0^1 \varphi\left(\int_0^1 a(u,t) \cdot b(t,v) dt\right) du dv$$

$$\leq \int_0^1 \int_0^1 \left(\int_0^1 a(u,t) \cdot \varphi(b(t,v)) dt\right) du dv$$

$$= \int_0^1 \int_0^1 \left(\int_0^1 a(u,t)du\right) \cdot \varphi(b(t,v))dtdv$$

$$= \int_0^1 \int_0^1 \varphi(b(t,v))dtdv = \tau(B) \tag{31}$$

where we have used Jensen's inequality and the following property

$$\int_0^1 a(u,t)du = \int_0^1 a(u,t)dt = 1 \tag{32}$$

The copula density based measures are in general symmetric measures, as noted in the examples above. Thus we also have $\tau(A * B) \leq \tau(A)$. As mentioned earlier, this is the issue with these symmetric measures: they take maximum value, which could be infinity for the above examples, on the set $\mathcal{L} * \mathcal{R}$ which includes a lot of noninvertible copulas.

The Sobolev norm based dependence measure as discussed in Siburg and Stoimenov (2009) is different as it does not satisfy either of the DPI conditions in Equation (23) although it is symmetric. The Sobolev norm based dependence measure is defined as

$$\tau_{Sob}^2(C) = 3 \int_0^1 \int_0^1 [(\partial_1(C(u,v) - \Pi(u,v)))^2 + (\partial_2(C(u,v) - \Pi(u,v)))^2]dudv \tag{33}$$

or

$$\tau_{Sob}^2(C) = \frac{1}{2}(\tau_2^2(C) + \tau_2^2(C^T)) \tag{34}$$

Thus

$$\tau_{Sob}^2(A * B) = \frac{1}{2}(\tau_2^2(A * B) + \tau_2^2(B^T * A^T)) \leq \frac{1}{2}(\tau_2^2(B) + \tau_2^2(A^T)) \tag{35}$$

which does not satisfy $\tau_{Sob}^2(A * B) \leq \tau_{Sob}^2(B)$ or $\tau_{Sob}^2(A * B) \leq \tau_{Sob}^2(A)$. This explains why the measure is only invariant under strictly monotonic transformations, but not under the general Borel-measurable bijections. It equals to one on the set of invertible copulas $\mathcal{L} \cap \mathcal{R}$ or copulas of mutual complete dependence.

The symmetric measures discussed by Schweizer and Wolff (1981) also do not satisfy DPI, thus are only invariant under strictly monotonic transformations.

### 5. Example

Here is an example (Nelson, 2006, Example 3.3) that demonstrates some of ideas discussed in the previous section.

Let $\theta \in [0,1]$ and define the copula as follows

$$C(u,v) = \begin{cases} u & if\ u \leq \theta v \\ \theta v & if\ \theta v < u < 1-(1-\theta)v \\ u+v-1 & if\ 1-(1-\theta)v \leq u \end{cases} \tag{36}$$

This copula is singular whose support consists of two line segments in $I^2$, one joining (0,0) and $(\theta,1)$, and the other joining $(\theta,1)$ and (1,0).

For this copula, $Y$ is completely dependent on $X$, but not the vice versa unless $\theta \in \{0,1\}$. This means that $C$ is left invertible $C^T * C = M$ but may not be right invertible $C * C^T \neq M$. It is easy to calculate the dependence measures $\tau_1(C) = 1$, $\tau_2(C) = 1$ and $\tau_1(C^T) = \theta^2 + (1-\theta)^2 \leq 1$, $\tau_2^2(C^T) = 3\left(\theta - \frac{1}{2}\right)^2 + \frac{1}{4} \leq 1$. Note that when $\theta = 0,1$, $C$ becomes also right invertible and $\tau_1(C^T) = \tau_2(C^T) = 1$; when $\theta = \frac{1}{2}$, $X$ is the least dependent on $Y$ and $\tau_1(C^T) = \tau_2(C^T) = \frac{1}{2}$. Again, we need both $\tau(C)$ and $\tau(C^T)$ to understand the dependence relation between $X$ and $Y$.

Let us look at the $*$ product between $C$ and a general copula $B$:

$$A = C * B = \int_0^1 \partial_2 C(u,t) \cdot \partial_1 B(t,v) dt$$

$$= \begin{cases} \int_0^{\frac{u}{\theta}} \theta \cdot \partial_1 B(t,v) dt & 0 \leq u \leq \theta \\ v + \int_0^{\frac{1-u}{1-\theta}} (\theta - 1) \cdot \partial_1 B(t,v) dt & \theta \leq u \leq 1 \end{cases} \tag{37}$$

Then

$$\partial_1 A = \begin{cases} \partial_1 B\left(\frac{u}{\theta}, v\right) & 0 \leq u \leq \theta \\ \partial_1 B\left(\frac{1-u}{1-\theta}, v\right) & \theta \leq u \leq 1 \end{cases} \tag{38}$$

Thus

$$\tau(A) = \int_0^1 \int_0^1 \varphi(\partial_1(A(u,v) - uv)) du dv$$

$$= \int_0^1 \left( \int_0^\theta \varphi\left(\partial_1 B\left(\frac{u}{\theta}, v\right) - v\right) du + \int_\theta^1 \varphi\left(\partial_1 B\left(\frac{1-u}{1-\theta}, v\right) - v\right) du \right) dv$$

$$= \int_0^1 \left( \theta \int_0^1 \varphi(\partial_1 B(u,v) - v) du + (1-\theta) \int_0^1 \varphi(\partial_1 B(u,v) - v) du \right) dv$$

$$= \int_0^1 \int_0^1 \varphi(\partial_1(B(u,v) - uv)) du dv = \tau(B) \tag{39}$$

We have assumed that the convex function $\varphi$ has no direct dependence on $u$ when we make a change of variable in the integral. So the $*$ product of the left invertible copula $C$ on the left will not change the dependence measure. In the special case where $B = C^T$ is right invertible, we notice that in general $\tau(C * C^T) = \tau(C^T) \leq 1$. This is different from the symmetric measure described in Ruankong, Santiwipanont and Sumetkijakan (2012), Corollary 5.5, where $C * C^T$ always has dependence measure 1.

For the right invertible copula, we can use the transpose $C^T$.

$$A = C^T * B = \int_0^1 \partial_1 C(t,u) \partial_1 B(t,v) dt$$
$$= \int_0^{\theta u} \partial_1 B(t,v) dt + \int_{1-(1-\theta)u}^1 \partial_1 B(t,v) dt \qquad (40)$$
$$= v - B(1-(1-\theta)u, v) + B(\theta u, v)$$

Then

$$\partial_1 A = \theta \cdot \partial_1 B(\theta u, v) + (1-\theta) \cdot \partial_1 B(1-(1-\theta)u, v) \qquad (41)$$

Thus

$$\tau(A) = \int_0^1 \int_0^1 \varphi(\theta \cdot \partial_1 B(\theta u, v) + (1-\theta) \cdot \partial_1 B(1-(1-\theta)u, v) - v)) du dv$$
$$\leq \int_0^1 \int_0^1 [\theta \cdot \varphi(\partial_1 B(\theta u, v) - v) + (1-\theta) \cdot \varphi(\partial_1 B(1-(1-\theta)u, v) - v)] du dv$$
$$= \int_0^1 \int_0^1 \varphi(\partial_1 B(u,v) - v)) du dv = \tau(B) \qquad (42)$$

where we have used Jensen's inequality again. So in general $\tau(A) \leq \tau(B)$.

## 6. Conclusion

We reviewed the original Rényi's axioms on symmetric dependence measures and proposed a new set of axioms that applies to nonsymmetric dependence measures. In the case of continuous random variables where the copula function exists uniquely, we showed that a new class of nonsymmetric nonparametric measures defined in terms of partial derivatives of copulas actually better characterize the relationship between a pair of random variable such that the measure takes maximum value only when one is completely dependent on the other. The measures also satisfy a Data Processing Inequality (DPI) on the $*$ product on copulas and thus lead to the satisfaction of the new axioms including the invariance of the dependence measure under bijective transformation on one of the random variables. The symmetric measures discussed in previous literature were shown to be inadequate to describe complete dependence.

Further research is needed to explore the usefulness of the new measures in various applications. A numerical estimation has already been proposed in Dette, Siburg and Stoimenov (2010) for regression dependence. It would be interesting to see that nonsymmetric measure can help us better understand dependence relations in vast amount of data.

# Appendix  Entropy-like nonsymmetric dependence measures

In this section, we extend the entropy-like symmetric dependence measures to nonsymmetric case. This includes Rényi's mutual information, Tsallis entropy and Shannon's mutual information.

The extension to Rényi's mutual information is as follows

$$R_\alpha^N(C) = \frac{1}{\alpha-1} \log \left( \int_0^1 \int_0^1 \left(\frac{\partial_1 C}{v}\right)^\alpha dudv \right) \qquad 0 < \alpha < 2 \qquad (A1)$$

Using Jensen's inequality,

$$R_\alpha^N(C) = \frac{1}{\alpha-1} \log \left( \int_0^1 \int_0^1 \frac{\partial_1 C}{v} \cdot \left(\frac{v}{\partial_1 C}\right)^{1-\alpha} dudv \right)$$

$$\geq \frac{1}{\alpha-1} \log \left( \int_0^1 \int_0^1 \frac{\partial_1 C}{v} \cdot \frac{v}{\partial_1 C} dudv \right)^{1-\alpha} = 0 \qquad (A2)$$

where we have used $\int_0^1 \int_0^1 \frac{\partial_1 C}{v} dudv = 1$. The lower bound holds when $\partial_1 C = v$ almost surely or $C$ is equivalent to the independent copula $\Pi$. Meanwhile, as $0 \leq \partial_1 C \leq 1$, $\partial_1 C^\alpha \leq \partial_1 C$ when $\alpha \geq 1$ and $\partial_1 C^\alpha \geq \partial_1 C$ when $0 < \alpha \leq 1$. So we have

$$R_\alpha^N(C) \leq \frac{1}{\alpha-1} \log \left( \int_0^1 \int_0^1 \frac{\partial_1 C}{v^\alpha} dudv \right) = \frac{1}{\alpha-1} \log \left( \int_0^1 v^{1-\alpha} dv \right) = -\frac{\log(2-\alpha)}{\alpha-1} \qquad (A3)$$

The upper bound holds when $\partial_1 C \in \{0,1\}$ almost surely or $C$ is left invertible. If $\alpha \geq 2$, $R_\alpha^N(C)$ is unbounded. Note that the original Rényi's mutual information is also not bounded. The new measure is bounded when $0 < \alpha < 2$ and is scaled with the constant $-\frac{\alpha-1}{\log(2-\alpha)}$ to the range [0,1]. It satisfies the DPI condition, so it is in the class of dependence measures discussed in the main article.

The extension to Tsallis entropy is as follows

$$\Delta_\alpha^N(C) = \frac{1}{\alpha-1} \left( \int_0^1 \int_0^1 \left(\frac{\partial_1 C}{v}\right)^\alpha dudv - 1 \right) \qquad 0 < \alpha < 2 \qquad (A4)$$

Similar argument leads to $\Delta_\alpha^N(C) \geq 0$ and $\Delta_\alpha^N(C) \leq \frac{1}{2-\alpha}$. So a scale constant of $2-\alpha$ makes $\Delta_\alpha^N(C)$ a dependence measure in [0,1]. Again, this new measure satisfies the DPI condition and most of the propositions in the main articles hold. The only exception is the proposition on the invariance under monotonic transformation on $Y$.

In the limit $\alpha \to 1$, $R_\alpha^N(C)$ and $\Delta_\alpha^N(C)$ both reduce to

$$R(C) = \int_0^1 \int_0^1 \frac{\partial_1 C}{v} \cdot \log \left(\frac{\partial_1 C}{v}\right) dudv \qquad (A5)$$

which is a nonsymmetric extension to Shannon's mutual information. Unlike the Shannon's mutual information which becomes infinity if the density of a copula has singularity, $R(C)$ is in the range $[0,1]$. Although Shannon's mutual information can be rescaled to $[0,1]$ through Linfoot's information coefficient of correlation, it equals to 1 for any copula with singularity, while $R(C) = 1$ only when $C$ is a copula of complete dependence.

A more general way to construct the dependence measures is as follows:

$$\hat{R}_\alpha^N(C) = \frac{1}{\alpha-1} \log \left( \int_0^1 \int_0^1 (k+2) v^{k+1-\alpha} \cdot \partial_1 C^\alpha \, du \, dv \right) \tag{A6}$$

$$\hat{\Delta}_\alpha^N(C) = \frac{1}{\alpha-1} \left( \int_0^1 \int_0^1 (k+2) v^{k+1-\alpha} \cdot \partial_1 C^\alpha \, du \, dv - 1 \right) \tag{A7}$$

where $0 < \alpha < k+3$ and $k > -2$, such that

$$\hat{R}_\alpha^N(C) = \frac{1}{\alpha-1} \log \left( \int_0^1 \int_0^1 (k+2) v^k \cdot \partial_1 C \cdot \left( \frac{v}{\partial_1 C} \right)^{1-\alpha} du \, dv \right)$$

$$\geq \frac{1}{\alpha-1} \log \left( \int_0^1 \int_0^1 (k+2) v^k \cdot \partial_1 C \cdot \frac{v}{\partial_1 C} \, du \, dv \right)^{1-\alpha} = 0 \tag{A8}$$

and

$$\hat{R}_\alpha^N(C) \leq \frac{1}{\alpha-1} \log \left( \int_0^1 \int_0^1 (k+2) v^{k+1-\alpha} \cdot \partial_1 C \, du \, dv \right)$$

$$= \frac{1}{\alpha-1} \log \left( \int_0^1 (k+2) v^{k+2-\alpha} \, dv \right) = \frac{1}{\alpha-1} \log \left( \frac{k+2}{k+3-\alpha} \right) \tag{A9}$$

where we have used

$$\int_0^1 \int_0^1 (k+2) v^k \cdot \partial_1 C \, du \, dv = 1 \tag{A10}$$

Similar argument leads to $\hat{\Delta}_\alpha^N(C) \geq 0$ and $\hat{\Delta}_\alpha^N(C) \leq \frac{1}{k+3-\alpha}$.

In the limit $\alpha \to 1$, we get

$$\hat{R}(C) = \int_0^1 \int_0^1 (k+2) v^k \cdot \partial_1 C \cdot \log \left( \frac{\partial_1 C}{v} \right) du \, dv \tag{A11}$$

such that $0 \leq \hat{R}(C) \leq \frac{1}{k+2}$.

Note that the general measures reduce to the special ones introduced first when $k = -1$.